\documentclass[a4paper,fleqn,usenatbib]{mnras}

\usepackage{newtxtext,newtxmath}

\usepackage[T1]{fontenc}
\usepackage{ae,aecompl}

\usepackage{graphicx}
\usepackage{tabu}

\newcommand{\boldforreferee}{} 

\newcommand*{\figuretitle}[1]{%
    {\centering
    \textbf{#1}
    \par\medskip}
}

\title[Finding bright $z \geq 6.6$ Lyman-$\alpha$ emitters with lensing]{Finding bright $z \geq 6.6$ Lyman-$\alpha$ emitters with lensing: prospects for \textit{Euclid}}

\author[Marchetti, Serjeant \& Vaccari]{L. Marchetti$^{1,2,3,4}$, S. Serjeant$^{1}$, M. Vaccari$^{3,4}$\\
$^1$School of Physical Sciences, The Open University, Milton Keynes, MK7 6AA, UK\\
$^2$South African Astronomical Observatory, Observatory Road, 7935, Observatory, Cape Town, South Africa\\
$^3$Department of Physics and Astronomy, University of the Western Cape, Robert Sobukwe Road, 7535 Bellville, Cape Town, South Africa\\
$^4$ INAF - Istituto di Radioastronomia, via Gobetti 101, 40129 Bologna, Italy}

\begin{document}

\date{Accepted 2017 June 20. Received 2017 June 15; in original form 2016 March 18.}

\pagerange{\pageref{firstpage}--\pageref{lastpage}} \pubyear{2017}

\maketitle

\label{firstpage}

\begin{abstract}
We model the $z \geq 6.6$ Ly$\alpha$ luminosity function to estimate the number of lensed high$-z$ Ly$\alpha$ emitters that may be detected by the {\boldforreferee \textit{Euclid} Deep Survey}. To span the whole range of possible {\boldforreferee predictions} we exploit two Ly$\alpha$ luminosity function models and two strong gravitational lensing models from the literature. We show that the planned {\boldforreferee \textit{Euclid} Deep Survey} observing 40\,deg$^2$ over the 920-1850 nm wavelength range down to a flux limit of $F_{lim}=5\times10^{-17}$\,erg\,s$^{-1}$\,cm$^{-2}$ will enable us to find between $\sim 0.85$ and $\sim 1.82$ deg$^{-2}$ lensed Ly$\alpha$ emitters at $z \geq 6.6$ depending on the adopted Ly$\alpha$ luminosity function and strong gravitational lensing model. The obvious [O{\sc ii}], [O{\sc iii}] and H$\beta$ contaminants of the Ly$\alpha$ lensed population will be identified with the help of {\it Euclid}'s spectral resolving power, while the SKA will enable the identification of the interloper population of H$\alpha$ emitters. By combining {\it Euclid} and the SKA, we will thus be able to identify, for the first time, a sample of $\sim 34$ to $\sim 73$ lensed Ly$\alpha$ emitters at $z \geq 6.6$.
\end{abstract}



\section{Introduction}\label{sec:introduction}

Over the last decade, interest has grown in the study of sources at very high redshift ($z>5$) thanks to the {\it Hubble Space Telescope} and its deep imaging and spectroscopic capabilities, which for the first time enable the study of galaxies close to the epoch of reionisation. {\boldforreferee In this context, the identification of Lyman Break Galaxies (LBGs) and Ly${\alpha}$ emitters (LAEs) at very high redshift is an important step to study galaxy formation in the early Universe. 
In fact, even though the detection of these very distant objects presents tremendous observational challenges, it is largely accepted that the young stellar populations hosted in these early galaxies should be responsible for the high energy photons (typically with $E > 13.6$ eV) that reionise the diffuse hydrogen in the intergalactic medium or IGM (Dressler et al. 2015).}
{\boldforreferee 
Thanks to deep near infrared imaging provided by {\it Hubble}'s WFC3, great progress has been made in such studies as many LBGs have been found above $z\sim6$. However, these studies have also proven that the observed population of bright LBGs ($L>L^*$) can only produce a small fraction of the flux required to balance the ionising budget, and that much larger numbers of fainter, unobserved galaxies would be needed (Bunker et al. 2010, Dressler et al. 2015).}

Constraining the number density or luminosity function (LF) of LAEs and its evolution with cosmic time is therefor crucial to quantify the Ly${\alpha}$ photon output during the epoch of reionisation. {\boldforreferee However,} the study of the LF of LAEs and its evolution at $z \sim 6$ and above is far from straightforward. Several studies show that the LAE LF appears to be constant between $z=3$ and $z\sim6$ (e.g., Gronwall et al. 2007, Ouchi et al. 2008), while at $z\geq6$, with the increase of neutral hydrogen (H{\sc i}) in the IGM, the number density of LAEs is expected to drop with redshift, but the rate of this drop is still {\boldforreferee rather unconstrained}. 

{\boldforreferee Recent studies, exploiting data from a number of deep surveys undertaken with e.g., the \textit{Subaru} telescope have brought to different conclusions on this matter. For example, Kashikawa et al. 2011 found a decrease of the LAE number density affecting at least the bright end (at $L\geq10^{42.5}$) of the LF between $z=5.7$  and $z=6.5$. Ouchi et al. 2010, Konno et al. 2014, 2017 also found that LAEs are dropping, following a negative density and luminosity evolution between $z=5.7$ and $z=7.3$. In particular, Konno et al. 2014 found an accelerated LF evolution above $z\sim6.5$. On the other hand, Matthee et al. 2014, 2015 and Santos et al. 2016 found a differential evolution of the LAE LF at $z>5$. They found that between $z=5.7$ and $z=6.6$ the number density of LAEs drops with negative pure density evolution at luminosities below $L\sim10^{43.5}$ erg s$^{-1}$, while at $L>10^{43.5}$ erg s$^{-1}$ the LF remains almost constant, suggesting no evolution. This trend seems to be confirmed also by the first photometric results of the Lyman Alpha Galaxies in the Epoch of Reionization (LAGER) survey presented in Zheng et al. 2017 and spectroscopically confirmed by Hu et al. 2017. Exploiting the Dark-Energy Camera (DECam) on the NOAO/CTIO 4m Blanco telescope LAGER identified 27 LAEs at $z=6.9$ whose luminosity function can be described by a Schechter function with an excess at $L\sim10^{43.4}$ erg s$^{-1}$ (Zheng et al. 2017). This excess is determined by only 4 galaxies so the result has to be taken with caution. At the same time though, Bagley et al. 2017, exploiting spectroscopic observation obtained as part of the WFC3 Infrared Spectroscopic Parallel (WISP, PI: M. Malkan, Atek et al. 2010) Survey, detected two LAEs at $z=6.5$ over $\sim0.04$ deg$^{-2}$ in agreement with the number density described by Matthee et al. 2015, but they also found no detections at $z=7$ or above, suggesting that the Ly${\alpha}$ luminosity function should evolve and drop above $z\sim6.6$ in a manner that seems to be in agreement with Faisst et al. 2014.}

These controversies can be in part explained by the numerous observational limitations affecting each of the mentioned surveys, such as the limited survey areas, the different flux limits, the effects of cosmic variance and the possible contamination by lower-redshifts interlopers. However, there may also be a more fundamental explanation linked to how reionisation takes place at $z>6$ and that we struggle to make sense of with current surveys. Some of the previous mentioned studies and other recent observational results suggest that at $z>6$ we can observe LAEs only if they are located in a ionised bubble which would allow the Ly${\alpha}$ photons to escape (e.g. Matthee et al. 2015, Stark et al. 2016, Castellano et al. 2016, Bagley et al. 2017, Zitrin et al. 2015, Zheng et al. 2017, Hu et al. 2017). In this interpretation, we would be able to preferentially observe the brightest LAEs at high redshift, as they would be the only objects able to ionise their surrounding and therefore to become visible. This trend would then be observed as a flattening of the bright end of the high-$z$ LAE LF. This model suggests that reionisation would not happen homogeneously across space, but that it would instead proceed differently depending on the luminosity and clustering properties of early galaxies. However, the limited data available and the large uncertainties on the H{\sc i} density distribution at the earliest epochs of galaxy formation make it difficult to discriminate between different scenarios. Only new deep and wide surveys searching for Ly${\alpha}$ emitters at high redshift will be able to solve this problem.

{\boldforreferee Strong gravitational lensing is a natural phenomenon that can be exploited to overcome some of the observational constraints described above and that, in some cases, could explain some of the recent findings. As an effect of gravitational lensing, surface brightness is conserved (due to Louiville's Theorem), but angular size can be increased, so the total flux can be magnified. Therefore, by definition, gravitational lensing helps to detect high-redshift objects otherwise too faint to be visible. These objects would thus appear brighter due to the lensing magnification, populating the brightest bins of the luminosity function and therefore changing its slope at the bright end.} In this work we investigate how the next generation of space missions such as {\it Euclid} will be uniquely placed to perform deep and wide spectroscopic surveys in which the steepness of the LF due to gravitational lensing will help to identify a population of very high-redshift LAEs. More specifically, we show that the planned {\boldforreferee \textit{Euclid} Deep Survey}, covering $\sim40$~deg$^2$ down to a fiducial flux limit of $F_{lim}=5\times10^{-17}$\,erg\,s$^{-1}$\,cm$^{-2}$) over the $920\leq\lambda\leq1850$ nm wavelength range spanned by \textit{Euclid}'s $blue$ and $red$ grisms (Costille et al. 2016), will be deep enough to observe the bright end of the luminosity function (LF) of a population of lensed LAEs at $6.5 \leq z \leq 9$. In the search of LAEs and strong lensing events, the identification of false positives due to contaminant populations is crucial and represents one of the biggest obstacle in surveys looking for very high-$z$ galaxy populations. In this work we show how we can eliminate contamination by other emission lines (such as H$\alpha$, [O{\sc ii}], [O{\sc iii}] and H$\beta$ emissions) thanks to {\it Euclid}'s spectral resolving power and its synergy with the Square Kilometre Array (SKA). We will then illustrate how we can use the steepness of the bright end of the observed Ly$\alpha$ luminosity functions to identify a population of $z\geq6.5$ LAE lensed candidates.

The paper is structured as follows: in section 2 we explain the method and the assumptions behind our selection method; in section 3 we show our results; in section 4 we discuss our estimates and in section 5 we summarise our conclusions and make some final remarks. For all the calculations reported in this paper we assume the following cosmology: Hubble constant of $H_0=70$\,km\,s$^{-1}$\,Mpc$^{-1}=100h$\,km\,s$^{-1}$\,Mpc$^{-1}$ and density parameters of $\Omega_{\rm M}=0.3$ and $\Omega_\Lambda=0.7$.
\section{Method}\label{sec:method}
In this work we exploit the very same successful strong gravitational lensing selection method used by Negrello et al. 2010 and Serjeant 2014. Very briefly, this method exploits the steepness of the luminosity functions (or number counts) as a tracer of the strong gravitational lensing magnification effect which is acting on a population of faint distant sources. 
Once the populations of contaminants are removed (e.g. local galaxies and blazars in the case of bright $500\,\mu$m emitters, Negrello et al. 2010), this method can guarantee a $\sim100\%$ reliability in selecting strong gravitational lensing events, as confirmed by a number of multi-wavelength spectroscopic and imaging follow-up campaigns of lens candidates identified in this way (e.g., Bussmann et al. 2013, Bussmann et al. 2015)

Following Negrello et al. 2010 and Gonz\'alez-Nuevo et al. 2012 who applied this selection method to the far-infrared luminosities of galaxies observed by {\it Herschel}, Serjeant 2014 showed that exploiting the strong magnification bias affecting the bright end of the H$\alpha$ luminosity function and of the H{\sc i} mass function we can identify up to $10^3$ and $10^5$ strong gravitational lenses with {\it Euclid} and the SKA respectively. The far-infrared, H$\alpha$ and H{\sc i} luminosities can thus facilitate the selection of strong gravitational lensing events and in this work we apply similar considerations to the Ly${\alpha}$ luminosity.

Over the last decade, as for the H$\alpha$ and far-infrared luminosities, the measurement of the redshifted Ly${\alpha}$ line has been used as a tracer of star formation. The Ly${\alpha}$ emission is linked to the photospheric emission of young stars and therefore is a direct tracer of the recent star formation rate (SFR). Moreover, this line is an order of magnitude stronger than the H$\alpha$ making it particularly efficient in tracing star formation at high redshift where it allows to detect both low-mass star-forming galaxies and intergalactic gas clouds (Ouchi et al. 2010, Kennicutt \& Evans 2012). Nevertheless, the use of the Ly${\alpha}$ emission as a quantitative SFR tracer not straightforward. This line is particularly sensitive to strong quenching from the combination of resonant trapping and eventual absorption by dust usually quantified as the Ly${\alpha}$ escape fraction, $f_{esc}$. Due to its resonant nature, the Ly${\alpha}$ transition causes the photons to scatter in the H{\sc i} present in the ISM, therefore, depending on the H{\sc i} characteristics such as its distribution and kinematics and on the dust content of a galaxy, the $f_{esc}$ value may easily vary between 0 and 1 (Haynes et al. 2010). For this reason, quantitatively translating the Ly${\alpha}$ luminosity in terms of SFR is highly uncertain, but Ly${\alpha}$ surveys can still be used for identifying large samples of distant star-forming galaxies especially at very high luminosities. As with the H$\alpha$ LF we can use the steepness of the Ly${\alpha}$ LF to identify strong gravitational lensing events.

However, while the H$\alpha$ LF is well suited to perform such a study (Serjeant 2014), observing Ly${\alpha}$ emitters especially at high redshift is still extremely challenging and therefore the statistics available to constrain the LAE LF is quite low. High uncertainties are associated with the shape of high-redshift LAE luminosity function and its evolution with redshift, and thus the interpretation of the results by the different authors is still rather controversial. With the aim of being as comprehensive as possible, and to bracket the most optimistic and most conservative likely scenarios, in our work we adopt two often-used Ly${\alpha}$ luminosity function estimates available in the literature: the Ly${\alpha}$ LF at $z=6.6$ estimated by Ouchi et al. 2010 and the Ly${\alpha}$ LF at $z=7.3$ by Konno et al. 2014. We then apply to each of these estimates the lensing formalism described below and we compare the results with others available in the literature. These estimates represent a good range of models for the evolution of the LAE luminosity function, with potentially different implications for reionisation. Moreover, they were both estimated using deep narrow-band photometric \textit{Subaru} \textit{Suprime-Cam} surveys combined with a plethora of multi-wavelength data. Despite the fact that the detection of the Ly$\alpha$ emitters below $L*\sim10^{43}\,\rm{erg}\,\rm{s}^{-1}$ seems to become rapidly incomplete at $z\geq5.7$ (Dressler et al. 2011 and Dressler et al. 2015), the wide field of view of \textit{Suprime-Cam} appears to be a very efficient way to find high redshift LAEs with luminosities $L>L^{*}$, making these estimates suitable to perform our predictions. 

The evolving luminosity functions adopted in this work are summarised hereafter, but for more details on their estimates we refer the reader to the original papers. Ouchi et al. 2010 estimated the Ly$\alpha$ luminosity function at $z=6.6$ using data from the Subaru XMM-Newton Deep Survey (or SXDS). They found that the Ly$\alpha$ luminosity function is described by a Schechter function with the following parameters: $\Phi^*= 8.5^{+3.0}_{-2.2}\times10^{-4}$ Mpc$^{-3}$, $L^*=4.4^{+0.6}_{-0.6}\times10^{42}$ erg s$^{-1}$ and a fixed $\alpha=-1.5$. They also found that the LF was not evolving much between $z=5.7$ and $z=6.6$ with a decrease of $30\%$ of the LF at all luminosities. Konno et al. 2014 estimated the Ly$\alpha$ luminosity function at $z=7.3$ exploiting deep narrow-band imaging in the SXDS and Cosmic Evolution Survey (COSMOS). They found an LF described by a Schechter function with $\Phi^*= 3.7^{+17.6}_{-3.3}\times10^{-4}$ Mpc$^{-3}$, $L^*=2.7^{+8.0}_{-1.2}\times10^{42}$ erg s$^{-1}$ and a fixed $\alpha=-1.5$. In particular they found an evolution more pronounced than in Ouchi et al. 2010 and described by a pure density evolution of $(1+z)^5$ between $z=5.7$ and $z=6.6$ and of $(1+z)^{20.8}$ at higher redshift.

In this paper we aim to predict the number of LAEs which may be detected by the {\boldforreferee \textit{Euclid} Deep Survey} covering the wavelength range 920-1850 nm, where the Ly$\alpha$ emission line ($\lambda_{rest} = 121.6 \,\rm{nm}$) would be visible if emitted by sources at redshifts $z>6.5$. To conduct our studies we thus estimate the LAE LFs in this redshift regime by evolving the Ouchi et al. 2010 and Konno et al. 2014 LFs with the pure density evolution described in each paper and mentioned above.
We then apply to each of these LFs the same formalism used in Serjeant 2014 to calculate the differential magnification probability distribution $p(\mu,z){\rm d}\mu$ as from Blain (1996), Perrotta et al. (2002) and Perrotta et al. (2003). In this formalism the high magnification tail has the form $p(\mu,z)=a(z)\mu^{-3}$ for a function $a(z)$ that varies according to the nature and evolving number density of lenses. We adopt the same two complementary approaches as in Serjeant 2014: in the case of Perrotta et al. (2002, 2003), the mass spectrum follows the Sheth and Tormen (1999) formalism and we assume that the galaxy lens population has a singular isothermal sphere profile; alternatively we also show the case of model "B" in Blain (1996) approach, where the mass spectrum of the lensing galaxies does not evolve with redshift. To compare the two models we normalise the obtained Blain (1996) $a(z)$ solution to the prediction of Perrotta et al. (2002, 2003) at an arbitrary redshift of $z=1$ (as we did in Serjeant 2014). For the maximum magnification imposed by the finite source sizes, we follow Perrotta et al. (2002). The range of plausible maximum magnifications is therefore between $\mu\leq10$ and $\mu\leq30$, as appropriate for source characteristic radii of $\sim1-10h^{-1}$\,kpc (e.g. Wuyts et al. 2013). This size constraint is very conservative as galaxies at very high redshifts appear to be physically small, with radii typically below $1h^{-1}$\,kpc and rapidly declining with increasing redshifts (Bouwens et al. 2006, Liu et al. 2016, Bouwens et al. 2016). The current {\boldforreferee understanding of} the sizes of very high-redshift galaxies is mainly based on HST observations, which is more sensitive to smaller high surface brightness objects than extended ones. Larger sources may exist that have not yet been detected by the existing surveys. If larger sources exist in the redshift range investigated in this work, these will simply increase the number of lensed sources and our predicted lensed surface density should therefore be considered a lower limit.

\begin{figure*}
\centering
\figuretitle{Ly$\alpha$ Luminosity Function Model: Ouchi et al. 2010}
\includegraphics[width=1.0\textwidth]{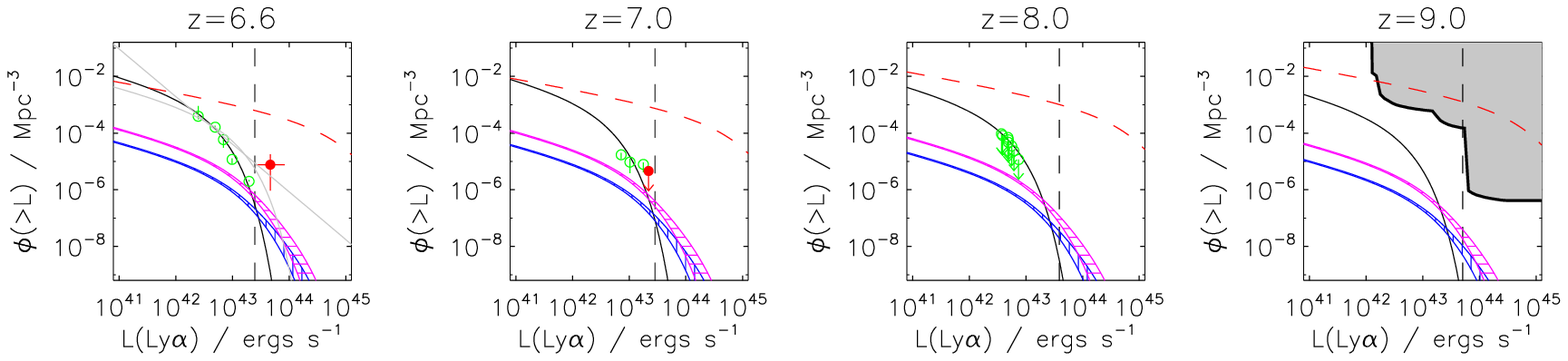}\\
\figuretitle{Ly$\alpha$ Luminosity Function Model: Konno et al. 2014}
\includegraphics[width=1.0\textwidth]{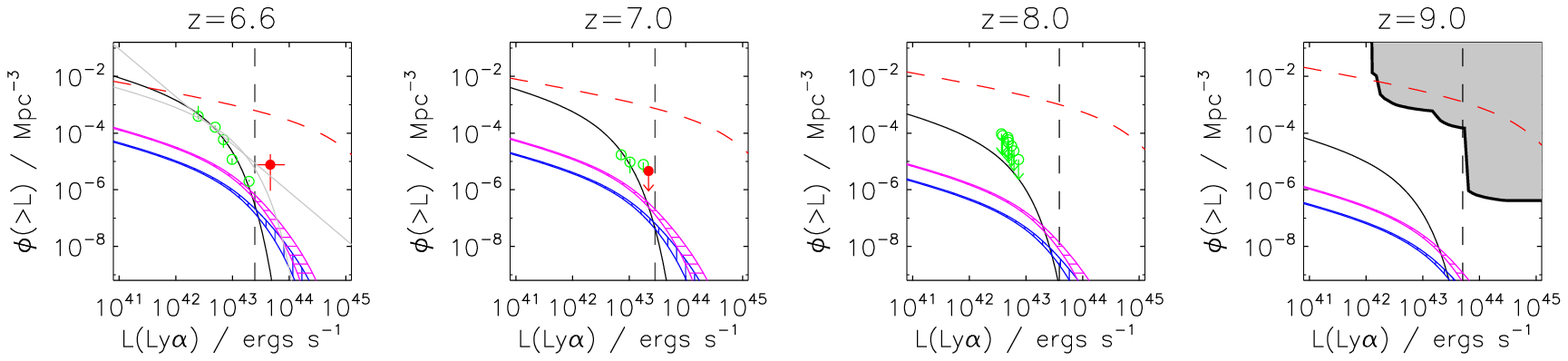}\\
\caption{\label{fig:lyalpha_lf}Ly$\alpha$ luminosity function model from Ouchi et al. 2010 (top panel, black line) and Konno et al. 2014 (bottom panel, black line) at $z=6.6, 7, 8, 9$. Lensed population predictions using singular isothermal sphere lenses of Perrotta et al. 2002 and Perrotta et al. 2003 are in pink; in blue is a non-evolving population of lenses normalised to the Perrotta predictions at an arbitrarily chosen $z=1$. The hatched regions span the range of maximum magnification $\mu\leq10$ and $\mu\leq30$. The vertical lines show the fiducial flux limit of the {\boldforreferee {\it Euclid} Deep Survey} as defined by the {\it Euclid} ``Red Book'' (Laureijs et al. 2011). The red dashed line shows the dominant LF of the H$\alpha$ contaminant population at $z=0.41, 0.48, 0.67$ and $0.85$. The grey lines in the panel $z=6.6$ are the Schecter function and power-law profile by Matthee et al. 2015. The red solid circles in the panel at $z=6.6$ and $z=7$ are the measured LF $z=6.5$ and the upper limit at $7<z<7.63$ respectively by Bagley et al. 2017. The green empty circles and arrows are the measured LF (at $z=6.6$ and $z=7$) and observational upper limits (at $z=7.7$ reported as reference in the panel at $z=8$) by Faisst et al. 2014. The grey region in the panel at $z=9$ outlines the observational constraints reported in Sobral et al. 2009. The green open circles and green arrows are respectively the observational measurements and upper limits reported by Faisst et al. 2014.
Once we eliminate the H$\alpha$ contaminant population with SKA observations, the population of Ly$\alpha$ emitters detected by {\it Euclid} is largely dominated by strong gravitational lens systems.}
\end{figure*}
\begin{figure*}
\centering
\includegraphics[width=0.47\textwidth]{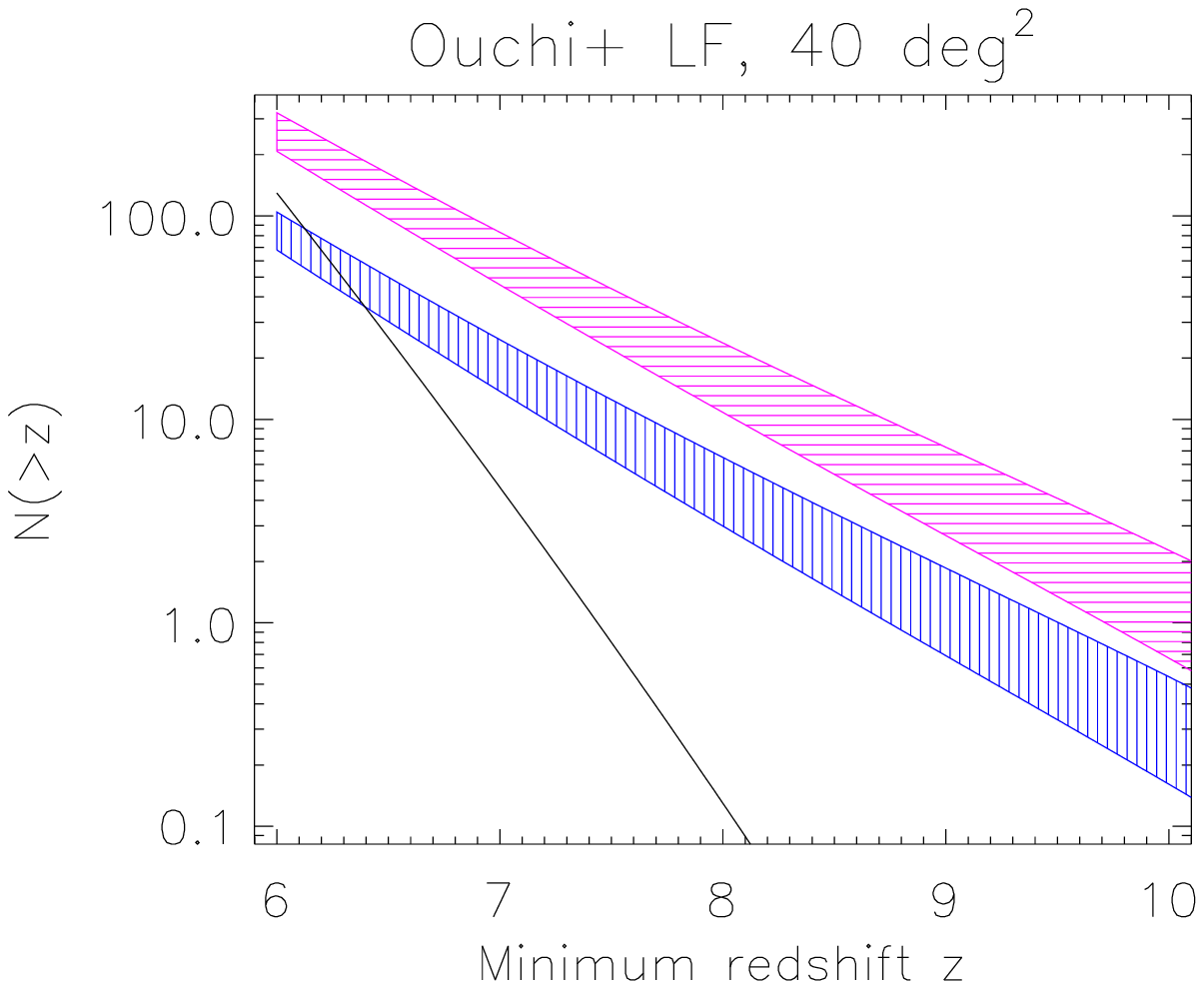}
\includegraphics[width=0.47\textwidth]{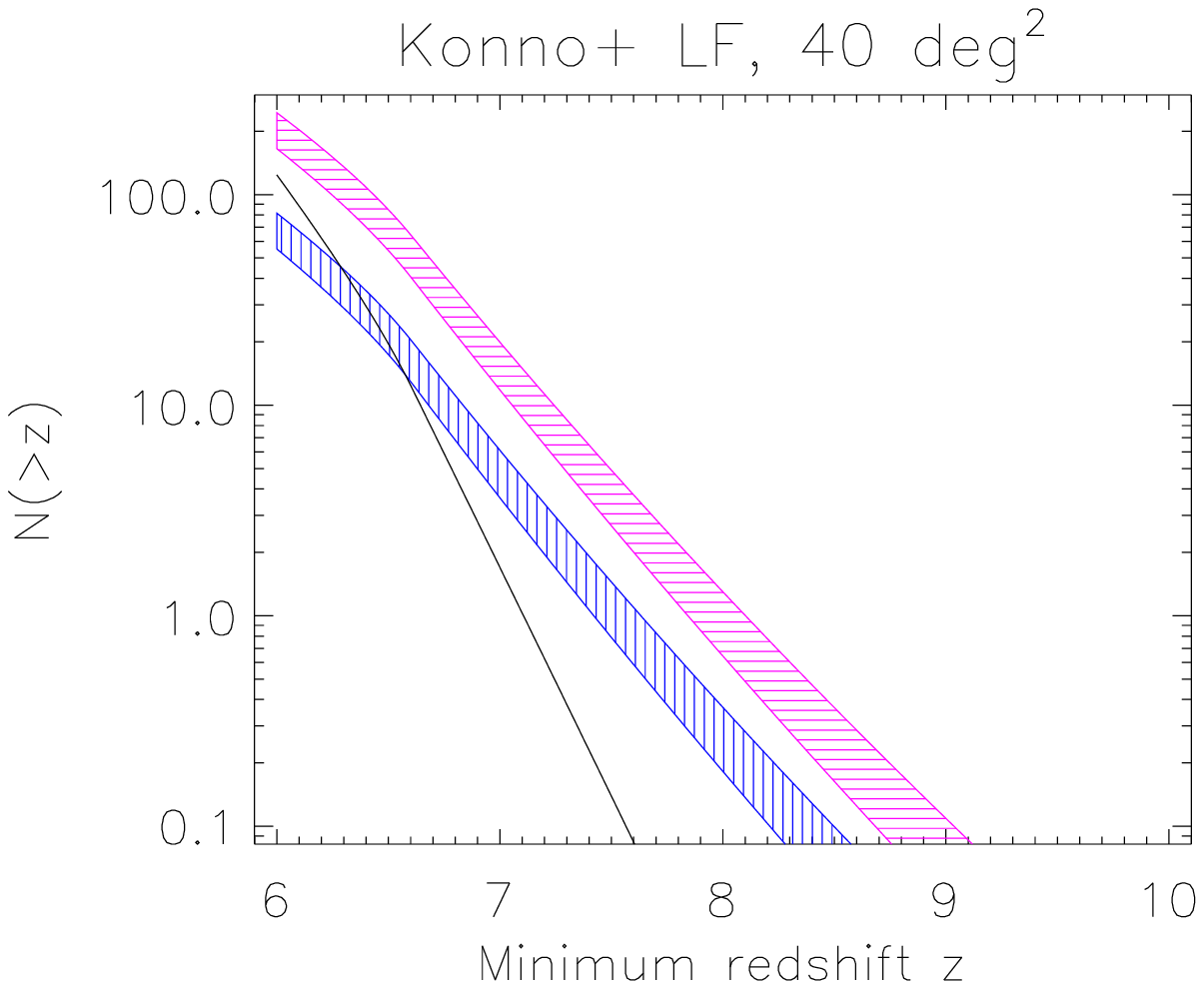}
\caption{\label{fig:lyalpha_counts}Cumulative Number counts for the unlensed and lensed population of Ly$\alpha$ emitters evolving with redshift and predicted using the models described in the text. The lensed population prediction are colour coded as in Fig.\,\ref{fig:lyalpha_lf}, while the unlensed population is shown with a solid black line.}
\end{figure*}

\section{Results}\label{sec:results}

In the wavelength range 920-1850 nm {\boldforreferee covered by the combination of \textit{Euclid}'s blue and red grisms}, the Ly$\alpha$ emission line would be visible if emitted by sources at redshfits $6.5<z<15$. However if we consider the fiducial flux limit of $5\times10^{-17}$\,erg\,s$^{-1}$\,cm$^{-2}$ envisaged for the {\boldforreferee {\it Euclid} Deep Survey} (Laureijs et al. 2011, reported with a black dashed vertical line in Fig.\,\ref{fig:lyalpha_lf}) at $z \geq 10$ even the magnified LAEs would be typically too faint to be detected. Moreover, if we adopt the Konno et al. 2014 evolutionary model, the drop of LAEs seems to be very accelerated at $z>7$, so that at $z \geq 9$ no LAEs would be visible (see also Sec.\,\ref{sec:discussion}). Therefore, in order to be able to perform a comparison between the different evolutionary models here adopted, we focus our attention on four redshift bins with mean redshifts $z=6.6, 7, 8, 9$ respectively. 

In Fig.\,\ref{fig:lyalpha_lf} we show the predicted Ly$\alpha$ luminosity function at $z=6.6, 7, 8, 9$ resulting by applying the lensing formalism described in the previous section to the Ly$\alpha$ LFs chosen for this study. Our predictions are also compared with empirical constraints from the literature. We found that, indipendently from the adopted LF model, our predictions are in agreement with the constraints reported in Sobral et al. 2009 and with the Ly$\alpha$ LF measurements and upper limits reported in Faisst et al. 2014 and Bagley et al. 2017. As a reference we also report the results at $z=6.6$ obtained by Matthee et al. 2015. While these results appear in agreement with our estimates at $z=6.6$, as already pointed out by Matthee et al. 2015, by assuming the Ouchi et al. 2010 and Konno et al. 2014 scenarios we might underestimate the number of sources in the brightest luminosity bins (see more discussion in Sec.\,\ref{sec:discussion}). This trend seems to be confirmed by Bagley et al. 2017 reported in red in Fig.\,\ref{fig:lyalpha_lf}. 

The same figure also illustrates the number density of H$\alpha$ contaminants whose identification is discussed in the next section. Once we eliminate H$\alpha$ and H$\beta$ lines as well as the [O{\sc ii}] and [O{\sc iii}] doublets, respectively at $\lambda_{rest} = 656.28,\, 486.1,\, 372.6,\, 372.9,\, 495.9,\, 500.7 \,\rm{nm}$ and all discussed in the next section, we notice that for both Ouchi's and Konno's LFs, at luminosities above $10^{44}$\,erg\,s$^{-1}$ at $6.6<z<9$, the observed population is dominated by strong gravitational lens systems, regardless of the maximum magnification or the nature of the lensing population. 

In Fig.\,\ref{fig:lyalpha_counts} we report the expected number of lensed and unlensed sources for each LF model adopted and as a function of redshifts assuming a fiducial sky coverage of $\sim40$\,deg$^2$ as for the envisaged \textit{ Euclid} Deep Survey. The expected number and density distribution of the predicted bright lensed galaxies at different redshifts for the different LF and lensing models are also reported in Tab.\,\ref{tab:counts}.

\begin{table}
\begin{small}
\begin{tabular}{l | c | c | c | c}
\hline
LF model & $z$ & \multicolumn{2}{c|}{\# lensed gal.} & $\Phi$  \\ 
                &       & (a) & (b) & [deg$^{-2}$]\\
\hline\hline
Ouchi 2010         & $z > 6.6$ & 26-43 & 83-140 & 1.82 \\
			   & $z > 7.0$ & 13-24 & 45-82 & 1.02 \\ 
		           & $z > 8.0$ & 3-6 & 10-23 & 0.26 \\ 
                            & $z > 8.6$ & 1-3 & 5-12 &  0.13 \\ 
\hline
Konno 2014         &  $z > 6.6$ & 13-20 & 40-64 & 0.85 \\ 
                             & $z > 7.0$ & 3-6 & 11-19 & 0.24 \\ 
                             & $z > 8.0$ & 0.17-0.34 & 0.6-1.2 & 0.01 \\ 
                             & $z > 8.6$ & 0.03-0.08 & 0.12-0.28 & 0.003 \\ 
\hline
\end{tabular}
\caption{\label{tab:counts} Expected number (over the $\sim40$\,deg$^2$ covered by the envisaged \textit{ Euclid} Deep Survey) and number density of the predicted bright ($L>12L^*$) lensed galaxies at different redshifts. The numbers are reported for each LF and lensing models considered in this study. The note (a) and (b) refers to the two lensing models adopted: (a) lensed population predictions using singular isothermal sphere lenses of Perrotta et al. 2002 and Perrotta et al. 2003 are; (b) non-evolving population of lenses normalised to the Perrotta predictions at an arbitrarily chosen $z=1$. The number separated by - refer to the two maximum magnification of 10 and 30 respectively and reported for each lensing model. The number density $\Phi$ [deg$^{-2}$] is estimated as the mean among the different lensing model predictions and magnifications.} 
\end{small}
\end{table}

%
\section{Discussion}\label{sec:discussion}
%
When modelling the Ly$\alpha$ LF and looking for high-redshift lensed galaxy candidates through their Ly$\alpha$ emission, the adopted LF model and the effective separation between lensed sources and contaminants are crucial.

The adopted high-redshift Ly$\alpha$ LF model must in particular be shown to be robust against the assumed high-redshift IGM transparency. Many authors have shown that the Ly${\alpha}$ seems to rapidly decline with increasing redshift, starting from $z>6$ (e.g. Stark et al. 2010, Schenker et al. 2012, Caruana et al. 2014, Treu et al. 2013, Konno et al. 2014, Furusawa et al. 2016, Ota et al. 2017), making its detection very uncertain. However, although this may be true at low luminosities (i.e. $M_{UV}>-22$), other recent work (e.g., Faisst et al. 2014, Matthee et al. 2015, Zitrin et al. 2015, Bagley et al. 2017, Zheng et al. 2017, Hu et al. 2017) has suggested a different scenario may apply to very luminous sources. For example, Faisst et al. 2014 (reported in Fig.\,\ref{fig:lyalpha_lf}) have been able to put some upper limits on the $z\sim7.7$ Ly${\alpha}$ LF faint end. Even though these are only upper limits and therefore cannot be used to make any strong claim on the real drop of the measured LF, they are consistent with our predicted bright end of the luminosity function at $z=8$. Others authors e.g., Barkana \& Loeb 2006, Matthee et al. 2014, 2015, Stark et al. 2016, Roberts-Borsani et al. 2016 showed that the disappearance of the LAE population may well be less pronounced in the most luminous galaxies in the reionisation era at least between $6<z<7$, suggesting new mechanisms in which the Ly${\alpha}$ would be able to escape efficiently even in this extreme environment. The framework suggested by these authors is at variance with the idea that reionisation would happen uniformly across the IGM. In this theory the reionisation is a {\boldforreferee patchier} process strongly linked with the luminosity and the clustering properties of the galaxies at play. In this case Ly${\alpha}$ photons emitted by the most luminous galaxies would be able to escape as the emitting source is ionising its environment creating an ionised bubble in an almost neutral IGM. More recently, Bagley et al. 2017 support this idea of ionised bubbles during reionisation, by detecting two LAEs at $z=6.5$ in WISP that verify Matthee et al. 2015 number density. However, they also claim that these two sources might not be able to ionise the bubble in which they reside on their own. They then suggest the presence of a bright quasar or a population of undetected low luminosities sources that would help in ionising the environment of their galaxies and that might be outside their field of view. The patchy resionisation scenario is also pictured by the more recent \textit{DECam} results by Zheng et al. 2017 and Hu et al. 2017. 
{\boldforreferee However, it must be noted that the samples analysed by Roberts-Borsani et al. 2016 and Stark et al. 2016 have been pre-selected to have [O{\sc iii}] and $H\beta$ emission with very high equivalent widths. This selection is biased to preferentially pick galaxies with young and massive stellar populations, which are also the more likely to show Ly${\alpha}$ emission. This has to be taken into account when we look at the results of these papers.}

In this work we show that even by assuming the {\boldforreferee most conservative scenario in terms of the high-redshift LAE LF estimated following Konno et al. 2014}, we still predict between 0.2 to 1.2 lensed LAE at $z>8$ in the envisaged \textit{Euclid} Deep Survey area of $\sim40\,$deg$^2$. There are plausible reasons for believing this is too conservative for the brightest populations. For example, Zitrin et al. 2015 reported the discovery in CANDELS of a Ly$\alpha$ emitter at $z=8.68$ with a flux between 1 and $2.5\times10^{-17}$ ergs/s/cm$^2$, i.e. between 2$\times$ and 5$\times$ fainter than the limit adopted for this study. Assuming the Konno et al 2014 LF, CANDELS should contain between 0.002 and 0.2 sources at these fluxes, depending on the flux of the Zitrin et al. 2015 object. Moreover, Konno's LF was poorly constrained at $L\geq10^{43}$ (as also remarked by Matthee et al. 2015), and this observational limit might be the cause of this discrepancy with other studies targeting brighter sources.

We can instead veer the other way and make a probably-too-optimistic prediction, by supposing that the earlier Ouchi et al. 2010 LFs estimated between $z=5.3$ and $z=6.6$ continue to higher redshifts. If this is the case, Ouchi's LF and evolution would predict between 0.04 and 6.8 sources in CANDELS in the brightness range of the Zitrin et al. 2015 object and this seems more in line with Zitrin's results. 

However, Ouchi's model is not the most optimistic LAE LF model presented in recent literature. For example, if the Matthee et al. 2015's evolutionary model were still true at $6.5<z<7$, Bagley et al. 2017 argue that we would be able to see $\sim$70 deg$^{-2}$ LAEs in the \textit{Euclid} Deep Survey we consider. As an exercise, we have applied the same lensing formalism described above to the power-law LF profile measured by Matthee et al. 2015 for sources with luminosity higher than $L\geq10^{43.5}$ at $z=6.6$. We have estimated how many lensed galaxies we would expect in the \textit{Euclid} Deep Survey, finding $\sim100$ and $\sim80$ lensed sources at $6.5\leq z\leq7$ applying to Matthee's LF the Ouchi's or Konno's evolutionary model respectively. These numbers have to be taken with great caution given the high uncertainties of the photometric data used by Matthee et al. 2015 to constrain the LF at $L\geq10^{43.5}$, and for this reason we have not included them as our main results. 

On the other hand, other studies of LBG luminosity functions have suggested an accelerated evolution at $z>8$ at variance with the evolution measured at $4<z<7$ (e.g. Bouwens et al. 2015) and these would then suggest that the abundance of the most luminous star forming galaxies may fall off very rapidly above $z=8$. These findings, even though they are based on the LBG population, might suggest that prediction at $z>7$ based on evolution at $6<z<7$ would underestimate this rapid drop in the counts of luminous galaxies also for the LAE population. If these findings are true, the solution pictured by Konno et al. 2014 would be the more accurate.

Turning to the problem of the most likely contaminants, for the redshift and wavelength range we are considering, these are the H$\beta$, [O{\sc iii}], [O{\sc ii}] and, most importantly, the H$\alpha$ interlopers (Dressler et al. 2015), although the possible AGN contamination must also be assessed. The [O{\sc iii}] emission doublet at $\lambda_{rest} = 495.9 \,\rm{nm}$ and $\lambda_{rest} = 500.7 \,\rm{nm}$ is not likely to be a problem as the near-infrared resolution of {\it Euclid} is $R\sim380$, higher than the $R=100$ required to resolve the [O{\sc ii}] doublet of the interloper population at $1.2<z<1.7$. The H$\beta$ ($\lambda_{rest} = 486.1 \,\rm{nm}$) emission is close in wavelength to [O{\sc iii}] so if we can detect the latter we can also eliminate the former in the same way. More critical is the [O{\sc ii}] emission doublet at $\lambda_{rest} = 372.6 \,\rm{nm}$ and $\lambda_{rest} = 372.9 \,\rm{nm}$, which would require a resolution $R>1200$ to be resolved. However, since [O{\sc ii}] interlopers would be at $2<z<2.6$, Euclid would be able to observe not only their 4000\,$\AA$ break but also their [O{\sc iii}] doublet, and the combination of broad-band photometry and spectroscopic line ratios would allow to effectively identify such interlopers, as e.g. done by Huang et al. 2016.

The potential H$\alpha$ contamination is more problematic. As shown in Fig.\,\ref{fig:lyalpha_lf}, the population of H$\alpha$ emitters that will be observed by Euclid in the same wavelength window as $6.6<z<9$ Ly$\alpha$ emitters is much more numerous than the population of Ly$\alpha$ emitters themselves. The apparent Ly$\alpha$ LF of the H$\alpha$ contaminants has been obtained by converting the H$\alpha$ LF used in Serjeant 2014 to an apparent Ly$\alpha$ LF using the following expression:
\begin{equation}
\Phi_{\rm{H}\alpha,cont}=\frac{\lambda_{rest,\rm{Ly}{\alpha}}}{\lambda_{rest,\rm{H}\alpha}}\,\frac{(\rm{dV}/{\rm{d}z})_{\rm{H}\alpha}}{(\rm{dV}/\rm{d}z)_{\rm{Ly}\alpha}}\,\Phi_{\rm{H}\alpha}
\end{equation}
where $\Phi_{\rm{H}\alpha,cont}$ is the LF of the H$\alpha$ contaminants (reported in red in Fig.\,\ref{fig:lyalpha_lf}), $\Phi_{\rm{H}\alpha}$ is the H$\alpha$ LF observed by Geach et al. 2010 while $\frac{\lambda_{rest\,\rm{Ly}{\alpha}}}{\lambda_{rest\,\rm{H}\alpha}}$ and $\frac{(\rm{dV}/{\rm{d}z})_{\rm{H}\alpha}}{(\rm{dV}/\rm{d}z)_{\rm{Ly}\alpha}}$ are the conversion factors required for the wavelength and the differential volume element respectively.

The contaminant H$\alpha$ population would dominate the Ly$\alpha$ population at all luminosities, therefore the identification of these interlopers is crucial in order to exploit our selection method. We estimate that the SKA will be able to detect such a population of H$\alpha$ interlopers as explained below.

The H$\alpha$ emission ($\lambda_{rest} = 656.28 \,\rm{nm}$) at $z=0.41, 0.48, 0.67$ and 0.85. overlaps with the Ly$\alpha$ emission at $z=6.6, 7, 8$ and 9 respectively. According to the {\it Euclid} "Red Book" (Laureijs et al. 2011), the fiducial flux limit for the {\it Euclid} Deep Survey is $5\times10^{-17}$\,erg\,s$^{-1}$\,cm$^{-2}$ which is indicated as a vertical line in Fig.\,\ref{fig:lyalpha_lf}.

This means that, in order to be detected by {\it Euclid}, the H$\alpha$ interloper population at lower redshifts need to be brighter than $2.49\times10^{40}$, $4.4\times10^{40}$, $9.6\times10^{40}$ and $1.8\times10^{41}$ erg s$^{-1}$ respectively.

\begin{figure}
\centering
\includegraphics[width=0.45\textwidth]{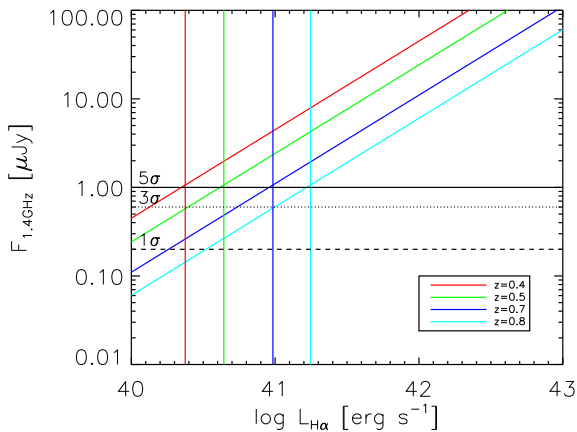} 
\caption{\label{fig:hae} {Representation of the expected radio 1.4 GHz fluxes of H$\alpha$ interlopers. Black lines: the SKA1 Deep Reference Survey flux limits at 1 (dashed line), 3 (dotted lines) and 5 $\sigma$ (solid line); colour-coded lines: H$\alpha$ luminosities corresponding to the {\it Euclid} Deep Survey spectroscopic flux limit (vertical lines) and as translated in radio 1.4 GHz (diagonal lines) following Kennicutt \& Evans 2012 at the different redshifts of the H$\alpha$ contaminant population considered in this work. Any H$\alpha$ interlopers seen by Euclid at the different redshifts will be detected by SKA1 at 5 $\sigma$ and above.}}
\end{figure}

If we correct these luminosities for dust extinction by 1 magnitude (as it is commonly-assumed by several authors e.g., Hopkins et al. 2004, Kennicutt et al. 2009; Pozzetti et al., 2016) and we convert them into SFRs and therefore into radio 1.4 GHz flux using the conversion factors by Kennicutt \& Evans 2012 we obtain $F_{1.4\rm{GHz}}\sim1.0 \mu$Jy (or a SFR of 0.3, 0.5, 1.3, 2.3 M$_{\odot} yr^{-1}$ respectively at $z=0.41, 0.48, 0.67$ and 0.85). In Fig. \ref{fig:hae} we illustrate the results of these conversions with respect to SKA1's fiducial flux limit.

The nominal noise level for the envisaged deep component of the SKA1 Reference Continuum Survey (see Prandoni \& Seymour et al. 2015) covering 10-30 deg$^2$ at 1.4 GHz is 0.2 $\mu$Jy/beam rms, which will allow us to identify the $H\alpha$ interlopers with a $S/N=5$. Therefore the Euclid Deep Spectroscopic Survey and the SKA1 are very well matched to carry out joint surveys in the equatorial regions and the southern hemisphere by the early 2020s. 
No radio facilities with capabilities similar to SKA1 are currently being planned in the Northern hemisphere, although the next generation VLA concept (Carilli et al. 2015) might in the future provide SKA1-like performance. The ongoing LOFAR Deep Surveys (Rottgering et al. 2006) will e.g. reach down to $\sim5\,\mu$Jy rms at 200 MHz. Since a $1\,\mu$Jy flux at 1.4 GHz corresponds to a 200 MHz flux of $\sim5\,\mu$Jy (using a spectral index $\alpha = 0.8$), LOFAR will only be able to detect {\it Euclid} H$\alpha$ emitters at $S/N=1$.

The SKA1 Deep Reference Continuum Survey will observe 10-30 deg$^2$ in the southern sky, most likely prioritising the LSST Deep Drilling Fields, i.e. currently COSMOS/ES1/XMM-LSS/CDFS, although additional deep fields are foreseen. However, due to its operational orbit, {\it Euclid} will only be able to carry out long integrations over sky areas that are situated at high ecliptic latitude, where visibility is highest. 
For this reason a more promising field for Euclid/SKA joint deep observations might then be the AKARI South Ecliptic Pole Field.
In this paper we report our estimates for the whole 40~$deg^2$ to be covered by the {\it Euclid} Deep Survey, but as detailed above in the foreseeable future we might only be able to identify the H$\alpha$ contaminants over the southern fields also observed by SKA1, i.e. possibly over 20 $deg^2$, which would scale down our predicted number of lenses by a factor of two, unless the next generation VLA offers comparable sensitivity to SKA1.

Moreover we need to check whether SKA would be able to directly detect the Ly$\alpha$ population at $z\geq8$. To implement this check we need to translate the typical luminosity of the population of LAE identified by {\it Euclid} into their expected radio fluxes using, as we do for the H$\alpha$, the relation between radio luminosity and SFR. Even though the conversion of the Ly$\alpha$ luminosity into SFR is highly uncertain, as already mentioned in Sec. \ref{sec:method}, in order to get at least a rough estimate of which SFR might be associated with our sample of $z\geq8$ LAEs we use the relation between the SFR(UV) and the SFR(Ly$\alpha$) suggested by Dijkstra et al. 2010. Combining this relation with the conversion reported in Kennicutt \& Evans 2012 we estimate that the typical radio flux of LAEs in the redshift range considered in our analysis would be a factor of $\sim10$ below the radio flux of the H$\alpha$ population and therefore negligible even considering the high uncertainties in the conversions applied. Moreover, the relation between radio luminosity and SFR may break down at ultra-high redshift, with the radio being strongly suppressed because of inverse Compton losses off the CMB (Lacki et al. 2010), so we can conclude that the SKA will be able to reliably discriminate between LAEs and H$\alpha$ emitters. Incidentally, modulo the uncertainties in the conversion between the [O{\sc ii}] luminosity and the SFR, the SKA1 Deep Reference Surveys would similarly allow us to also rule out any [O{\sc ii}] interlopers.

Finally, one must consider the possible contamination due to AGN line emission. However we believe that we can identify most of lower-z AGN by mean of their [O{\sc iii}] resolved emission. {\boldforreferee Moreover, [N{\sc ii}] ($\lambda_{rest}$=658.4nm) emission is strong in AGNs and will be marginally resolved from H$\alpha$ in \textit{Euclid} spectra, and AGNs will thus show up like unusually broad H$\alpha$ emitters or strong [N{\sc ii}] emitters.} On the other hand, many efforts are currently underway to establish reliable methods to distinguish star forming galaxies and AGNs in the ultra high-redshift regime, although most of them will require high spectral resolution follow-up. Konno et al. 2016 pointed out that all the LAEs at $z>2.2$ at $L>10^{43.4}$ may be AGNs.  However, Bagley et al. 2017, for example, have only been able to rule out the possibility that the sources observed by WISP were broad-line AGN, but the limits of their observation did not allow them to distinguish between narrow-line AGNs and star forming galaxies. This type of studies will become more straightforward  once JWST and ELTs will start operations (see e.g. Baek et al. 2013, Afonso et al. 2015, Feltre et al. 2016). 
\section{Conclusions}\label{sec:conclusions}
In this work we have shown that even applying the most conservative evolutionary model for the LAE luminosity function, the {\it Euclid} Deep Survey ($F_{lim}=5\times10^{-17}$\,erg\,s$^{-1}$\,cm$^{-2}$) will be able detect between $\sim$0.85\,deg$^{-2}$ and $\sim$1.82\,deg$^{-2}$ lensed Ly$\alpha$ emitters at $z\geq6.6$ according to the different LF and lensing model adopted. Our predictions were made considering a $\sim$40\,deg$^{2}$ envisaged footprint and a wavelength coverage between 920 and 1850 nm. Over this area and redshift range we will thus be able to observe a sample of between $\sim34$ and $\sim73$ strong gravitational lenses with known source redshifts, high signal-to-noise and identified by mean of a magnification-based selection independent of assumptions on lens morphology. 
This selection method will therefore substantially increase the number of known very high-redshift galaxies greatly contributing towards cosmological studies of the epoch of reionisation and providing excellent candidates for follow-up observations with JWST and ELTs. Nevertheless, only combining {\it Euclid} with ultra-deep radio continuum observations such as those foreseen by the SKA1 Deep Reference Survey we will be able to distinguish the population of Ly$\alpha$ lensed galaxies from the more numerous H$\alpha$ contaminants at lower redshifts.

\section*{Acknowledgements}

The authors thank the anonymous referee for helpful suggestions. LM and SS thank the Science and Technology Facilities Council for support under grant ST/J001597/1 and many colleagues in the {\it Euclid} Strong Lensing and Galaxy Evolution working groups for stimulating discussions.  LM acknowledges support from a South African DST-NRF Fellowship for Early Career Researchers from the United Kingdom, and thanks the South African Astronomical Observatory for their kind hospitality while this paper was being written. MV and LM acknowledge support from the EC REA (FP7-SPACE-2013-1 GA 607254 - Herschel Extragalactic Legacy Project), the South African DST (DST/CON 0134/2014) and the Italian MAECI (PGR GA ZA14GR02 - Mapping the Universe on the Pathway to SKA).

%


%
\bsp
\label{lastpage}
\end{document}